\documentclass[final,english]{bullsrsl}[2022/06/15]

\usepackage[latin1]{inputenc}
\usepackage[T1]{fontenc}
\usepackage{natbib} 
\usepackage{graphicx}

\begin{document}

\title{Necessity of a TDI optical corrector for ILMT observations}

\author[affil={1,2}, corresponding]{Vibhore}{Negi}
\author[affil={1,2}]{Bhavya}{Ailawadhi}
\author[affil={3,4}]{Talat}{Akhunov}
\author[affil={5}]{Ermanno}{Borra}
\author[affil={1,6}]{Monalisa}{Dubey}
\author[affil={1,6}]{Naveen}{Dukiya}
\author[affil={7}]{Jiuyang}{Fu}
\author[affil={7}]{Baldeep}{Grewal}
\author[affil={7}]{Paul}{Hickson}
\author[affil={1}]{Brajesh}{Kumar}
\author[affil={1}]{Kuntal}{Misra}
\author[affil={1,8}]{Kumar}{Pranshu}
\author[affil={7}]{Ethen}{Sun}
\author[affil={9,10}]{Jean}{Surdej}
\affiliation[1]{Aryabhatta Research Institute of observational sciencES (ARIES), Manora Peak, Nainital, 263001, India}
\affiliation[2]{Department of Physics, Deen Dayal Upadhyaya Gorakhpur University, Gorakhpur, 273009, India}
\affiliation[3]{National University of Uzbekistan, Department of Astronomy and Astrophysics, 100174 Tashkent, Uzbekistan}
\affiliation[4]{ Ulugh Beg Astronomical Institute of the Uzbek Academy of Sciences, Astronomicheskaya 33, 100052 Tashkent, Uzbekistan}
\affiliation[5]{Department of Physics, Universit\'{e} Laval, 2325, rue de l'Universit\'{e}, Qu\'{e}bec, G1V 0A6, Canada}
\affiliation[6]{Department of Applied Physics, Mahatma Jyotiba Phule Rohilkhand University, Bareilly, 243006, India}
\affiliation[7]{Department of Physics and Astronomy, University of British Columbia, 6224 Agricultural Road, Vancouver, BC V6T 1Z1, Canada}
\affiliation[8]{Department of Applied Optics and Photonics, University of Calcutta, Kolkata, 700106, India}
\affiliation[9]{Institute of Astrophysics and Geophysics, University of Li\`{e}ge, All\'{e}e du 6 Ao$\hat{\rm u}$t 19c, 4000 Li\`{e}ge, Belgium}

\correspondance{vibhore@aries.res.in}
\date{13th May 2023}
\maketitle

\begin{abstract}
The International Liquid Mirror Telescope (ILMT) has recently become operational at the Devasthal Observatory of ARIES, Nainital, India. The ILMT observes in the Time delay integration (TDI) mode where the images are formed by electronically stepping the charges over the pixels of the CCD, along a column. Observations near the zenith impose certain constraints dependent on the latitude such as image deformation due to the star-trail curvature and differential speed. These effects make the stellar trajectories in the focal plane of the ILMT to be hyperbolic, which are corrected for by the introduction of a TDI optical corrector, designed specifically for the ILMT. Here, we report the first results on the effect of this corrector on the trajectories followed by the stars in the ILMT focal plane. Astrometrically calibrating nine nights of data recorded with the ILMT during its first  commissioning phase, we find simple (nearly linear) relations between the CCD-y coordinate and the right ascension (RA) of stars and between the CCD-x coordinate and their  declination (DEC), respectively, which confirms that the TDI corrector works very fine in converting the stellar trajectories into straight lines.
\end{abstract}


\section{Introduction}
The 4m International Liquid Mirror Telescope (ILMT) is right now in its commissioning phase in the central Himalayan region near Nainital, India (79$^{\circ}$41$'$07.08$''$ E, 29$^{\circ}$21$'$41.4$''$ N, \citealt{2022JAI....1140003K}). The ILMT is based upon the liquid mirror technology where the primary mirror is a highly reflecting liquid (mercury) which takes the shape of a paraboloid when spun and images the celestial objects in its focal plane \citep{2018BSRSL..87...68S}. The ILMT is equipped with a 4K $\times$ 4K CCD camera, which is capable of observing in the {\it g}, {\it r}, and {\it i} bands and offers a field of view (FoV) of 22$'$ by 22$'$ \citep{2022JAI....1140001D}.
However, as the ILMT is non steerable, it cannot track the sources transiting in the sky, and to do the integration of photons coming from the sources, a special technique known as Time Delay Integration (TDI, also known as drift scanning) mode is used. In the TDI mode, images are formed by electronically stepping the relevant charges collected in the pixels of the CCD used as a detector \citep{1980SPIE..264...20M,1984MNRAS.210..979H}. The transfer rate is kept similar as the target drifts across the detector (i.e. equal to the observatory sidereal rate). \par
The TDI mode has several advantages over conventional observing modes, the most important being the required 1-D flat field correction \citep[][hereafter HR98]{1998PASP..110.1081H}. However, the images obtained in the TDI mode suffer from several deformations such as a declination dependent curvature of the star trails and a differential speed effect, that may result in significant loss of resolution with modern large-format CCD arrays (HR98, also \citealt{2002A&A...388..712V}). To account for these deformations, a curvature compensated TDI corrector has been proposed by HR98 and is currently being used at the prime focus of the ILMT facility, designed by \href{https://www.optisurf.com}{Optical Surfaces Ltd}. It  is to be noted that although several telescopes have worked/ been working in the TDI mode, this is the first instance of a TDI optical corrector being used in a telescope. Here, we report the first results on the stellar trajectories followed in the focal plane of the ILMT after introduction of the TDI optical corrector and also present some results on the astrometry of the ILMT data. 
The present paper is summarized as follows. In Section\,2, we discuss the possible deformations in the TDI mode observations followed by a briefly description of the TDI optical corrector in Section\,3. In Section\,4, we discuss the data used and methodology followed by some results in Section\,5, obtained for the astrometry of the detected objects with the ILMT. We present a brief discussion of our results in Section\,6 and summarize our conclusions in Section\,7.

\section{Deformations in the TDI imaging mode}
As discussed in Section 1, the stellar images acquired in the TDI mode suffer from certain deformations, especially for a non zero latitude, which may degrade the image quality and result in significant loss in the resolution. Below we briefly discuss some of these possible deformations.

\subsection{Star-trail curvature effect}

The CCD at the prime focus of the ILMT works in the TDI mode, where electrons are transferred from row to row along a rectilinear trajectory, at the same speed as the stars transit across the CCD.  However, the trajectory of a star with declination $\delta$ is not rectilinear, which is represented by the following equations (see \citealt{1996AJ....111.1721S}):
\begin{equation}
    \xi = f~\frac{\sin H ~\cos\delta}{\sin\delta_{0} ~\sin\delta + \cos H~ \cos\delta_{0}~\cos\delta}
\label{eq:eq1}
\end{equation}
\begin{equation}
    \eta = f~\frac{\cos H~\sin\delta_{0}~\cos\delta -\cos\delta_{0}~\sin\delta}{\sin\delta_{0}~\sin\delta + \cos H ~\cos\delta_{0}~\cos\delta}
\label{eq:eq2}
\end{equation}
where $\xi$ and $\eta$ are the positions along the East-West and North-South axis respectively, in a reference frame centred on the optical centre O, $\delta_{0}$ is the declination of the optical axis of the telescope (i.e., equal to the latitude of the telescope site), f is the focal length of the telescope, and H is the hour angle.
As can be seen, Equations~\ref{eq:eq1} and \ref{eq:eq2} are the equations of a conic section, which can be either an ellipse, a parabola or a hyperbola depending on whether the declination $\delta$ is greater than, equal to, or less than the co-angle ($\pi - \delta_{0}$, or the co-latitude for a zenith pointing telescope), and hence the trajectories followed by the stars in the focal plane are not linear, and result in an image elongation along the North-South axis. For a pictorial representation of the Star-trail curvature effect, please see Fig.\,1 of HR98 and Fig.\,4 of \citet{SurdejJProceedings1}.

\subsection{The differential speed effect}

In addition to the curvature effect, for a stellar trajectory which is non-central, there is a speed variation effect, i.e., the star will have a linear velocity smaller/larger than the central one as it enters the CCD before and exits after the latter one or vice versa. This effect will be worst for a star transiting at the top/bottom edge of the CCD, and causes an image elongation along the East-West direction.

\section{TDI corrector}

The image degradation discussed in the above section increases with the angular FoV and its correction is very crucial for a large FoV facility like the ILMT. A solution to the curvature effect was proposed by HR98 to use an optical corrector that eliminates the image displacement due to the curvature effect and the drift rate variation. The proposed corrector lens deploys decentered and tilted elements which can be varied to be used with any zenith pointing telescope at any latitude upto at least $\pm$50 degree. One such optical corrector has been installed at the prime focus of the ILMT, which is a combination of 5 lenses which are slightly offset and tilted to introduce an asymmetric field distortion that compensates for the nonlinear motion of the images. The ILMT corrector lenses are frozen in the structure for a latitude of $\pm$ 29$^{\circ}$22$''$, i.e., including the latitude of the Devasthal observatory. With the introduction of this optical corrector at the prime focus of the ILMT, the images are expected to move along parallel straight lines at a constant rate.

\section{Data Used and Methodology}
The data set used in this work was obtained with the ILMT during the 23 Oct - 01 Nov 2022 period. The data was pre-processed by correcting for dark, flat and cosmic removal. Each pre-processed TDI image consists of 4096 pixels (along DEC) $\times$ 36864 pixels (along RA), and covers 22$'$ in DEC and 3.3$^{\circ}$ in RA. The data was then astrometrically calibrated (Negi et al., in preparation, for the astrometric calibration pipeline).
In brief, all the sources within each frame were detected and registered and a subset of these objects was identified using the plate solving engine `astrometry.net' \citep{2010AJ....139.1782L}. After this, the entire field is crossmatched with the GAIA catalog making use of the precise astrometry of the GAIA survey. Once the precise equatorial coordinates from GAIA are known along with the CCD pixel positions of the point-like sources detected on the CCD, we fit  transformation relations between the equatorial coordinates and the pixel coordinates of all the stars. In case the star trajectories in the ILMT focal plane would have been hyperbolic, the transformation relations between the CCD-pixel coordinates and the stars' equatorial coordinates would have been rather complex, with terms included for the trajectory curvature. However, due to the correction by the TDI optical corrector, we expect simple transformation relations between the CCD-pixel coordinates and the equatorial coordinates. For this, we first converted the J2000 GAIA coordinates of all the stars to the observation epoch making correction for the precession, nutation of the Earth and light aberration. As the next step, we fit simple linear transformation equations (see Equations~\ref{eq:eq3} and \ref{eq:eq4}) between the CCD-pixel to equatorial coordinates (observation epoch). It is to be noted that in order to minimize possible errors due to non negligible proper motion and parallax effects, we use only stars with a proper motion < 20 mas/yr and a  parallax < 10 mas (see \citealt{2022JAI....1140001D} for details) while fitting the transformation equations
\begin{equation}
\alpha = f1 + [(x-x_{0}) * f2] + [(y-y_{0}) * f3]
\label{eq:eq3}
\end{equation}
\begin{equation}
\delta = g1 + [(x-x_{0}) * g2] + [(x-x_{0})*(x-x_{0})*g3]
\label{eq:eq4}
\end{equation}
where $\alpha$ and $\delta$ are the equatorial coordinates of the stars, {\it x} and {\it y} are their CCD pixel positions, {\it x$_{0}$, y$_{0}$} being the central CCD coordinates, and {\it f1, f2, f3, g1, g2} and {\it g3} being free parameters. \par

\section{Results}

\begin{figure}[t]
\centering
\includegraphics[width=0.99\textwidth,height=0.35\textheight]{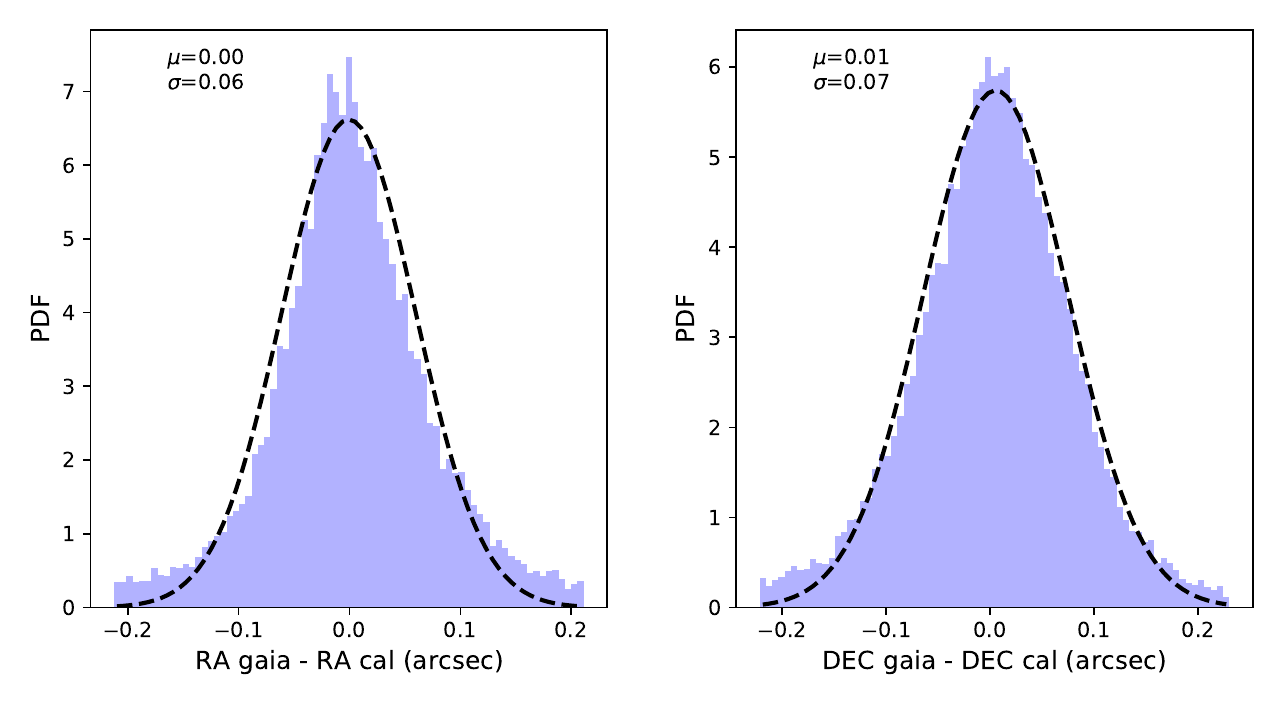}
\begin{minipage}{12cm}
\caption{{\it Left panel:} Distribution in the differences (calibrated - GAIA) of RA calculated using the transformation Equations~\ref{eq:eq3} and \ref{eq:eq4}, from a TDI frame observed with the ILMT on 25-10-2022. The dashed curve shows the Gaussian function fitted to the distribution with the best fit parameters shown in top left of each panel. {\it Right panel:} Same as in Left panel but for DEC.}
\label{fig:offset_hist}
\end{minipage}
\end{figure}

\begin{figure}[t]
\centering
\includegraphics[width=0.99\textwidth,height=0.3\textheight]{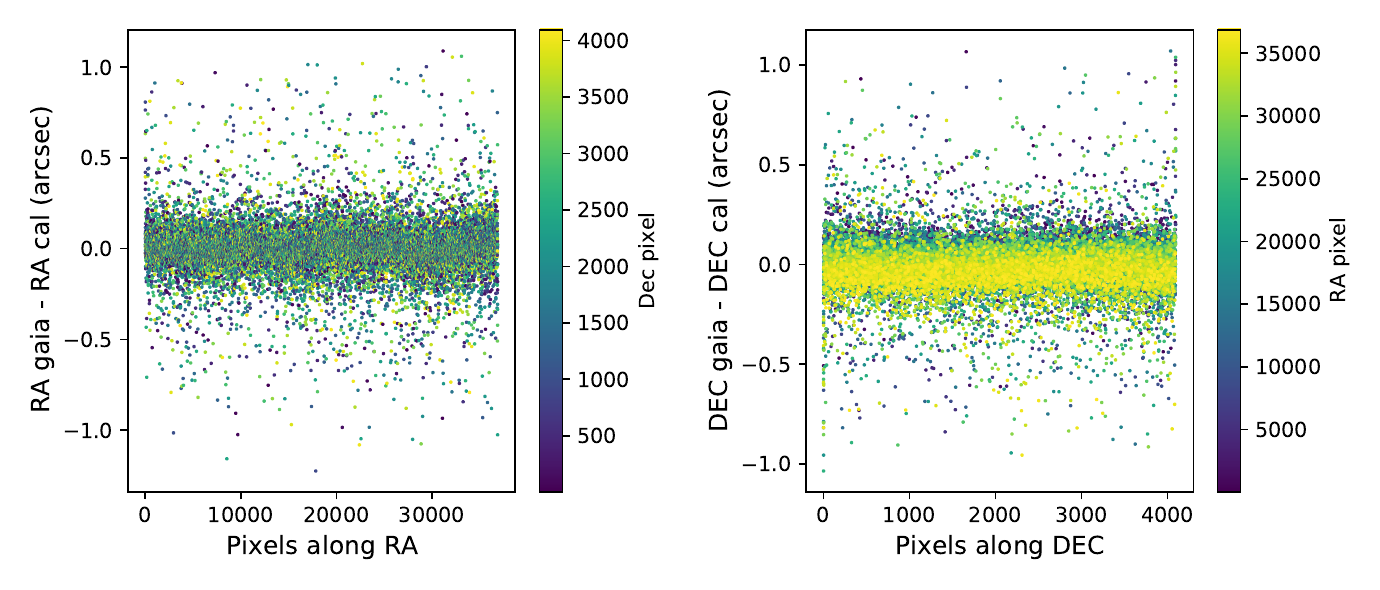}
\begin{minipage}{12cm}
\caption{{\it Left panel:} Difference (calibrated - GAIA) of RA calculated using the transformation Equations~\ref{eq:eq3} and \ref{eq:eq4}, from a TDI frame observed with the ILMT on 25-10-2022. The pixels are color coded with pixels along the DEC. {\it Right panel:} Same as in Left panel but for DEC.}
\label{fig:offset_scatter}
\end{minipage}
\end{figure}

\begin{table}
\centering
\caption{The best fit parameters obtained from Equations~\ref{eq:eq3} and \ref{eq:eq4} for one of the TDI frames obtained with the ILMT on 25 Oct 2022.}
\label{tab:fit_params}
\bigskip

\setlength\tabcolsep{4.5pt}
\begin{tabular}{cccccc}
\hline
\textbf{f1} & \textbf{f2} & \textbf{f3} & \textbf{g1} & \textbf{g2} & \textbf{g3}\\
\hline
74.620729 & -1.055946$\times10^{-8}$ & 0.000104 & 29.352263 & -9.082793$\times10^{-5}$ & 3.789334$\times10^{-11}$ \\
\hline
\end{tabular}
\end{table}
Using the transformation relations shown in Equations~\ref{eq:eq3} and \ref{eq:eq4}, the astrometric calibration of the 9 nights of data obtained with the ILMT, resulted in an astrometric accuracy of 0.05$''$ - 0.1$''$.
Fig.\,\ref{fig:offset_hist} shows the distribution of the differences (calibrated - GAIA) in the RA and DEC of one of the TDI frames obtained with the ILMT on 25-Oct 2022, where an astrometric accuracy of 0.06$''$ and 0.07$''$ has been  achieved for RA and DEC, respectively. Similarly, Fig.\,\ref{fig:offset_scatter} shows the differences (calibrated - GAIA) in RA and DEC, respectively. It can be seen that no systematic trend is present along any of these directions. The best fit parameters from Equations~\ref{eq:eq3} and \ref{eq:eq4}, for the same TDI frame are shown in Table\,\ref{tab:fit_params}. The astrometric accuracy which is better than one tenth of an arcsec confirms that the two Equations~\ref{eq:eq3} and \ref{eq:eq4} are the best transformation relations between the equatorial coordinates and the pixel coordinates, and the trajectories followed by the stars in the ILMT focal plane are indeed straight lines just affected by a very tiny rotation of the axes. Note the very small value of the $g3$ quadratic parameter in Table\,\ref{tab:fit_params}.

\section{Discussion}
Thanks to the very good performance of the TDI optical corrector, the stellar trajectories  in the focal plane of the ILMT should be straight lines. In that case, the RA of the star should be a function of the CCD y coordinate only and the DEC should be a function of the CCD-x coordinate only. However, in the case where there is a slight mis-alignment of the CCD rotation angle with respect to the East-West and South-North axes, RA should also show a dependence on the CCD-x coordinate. As has been found in our analysis, the best transformation relations between the stellar equatorial coordinates and their CCD-pixel ones, are presented by Equations~\ref{eq:eq3} and \ref{eq:eq4}.
As can be seen in Table\,\ref{tab:fit_params}, the parameter `f2' is very small whereas `f3' is relatively large, which shows that the RA has a linear dependence on the CCD-y coordinate whereas it has a very small dependence on the CCD-x coordinate, due to a slight misalignment of the CCD rotation angle. Here, the parameter `f3' corresponds to the plate scale (9.0833$\times10^{-5}$ deg per pixel) of the CCD. 
Similarly, from Equation~\ref{eq:eq4} and Table\,\ref{tab:fit_params}, it can be seen that the DEC has no dependence on the CCD-y coordinate, and has a linear dependence on the CCD-x coordinate only, where the parameter `g2' corresponds to the plate scale of the CCD along the South-North direction, along with a very small quadratic term possibly arising from the asymmetric TDI corrector where some lenses are slightly tilted and/or offset from optical axis. These relations confirm that the TDI optical corrector works very well in converting the stellar trajectories in the focal plane of the ILMT into straight lines. \par

\section{Conclusion}
In this work, we have checked the effect of the TDI optical corrector on the trajectories followed by the stars in the ILMT focal plane. Using the data obtained with the ILMT during its  first commissioning phase, and calibrating it astrometrically, we have achieved an astrometric accuracy of 0.05$''$ - 0.10$''$, using transformation relations that have a nearly linear dependence of RA versus the CCD-y coordinate, and DEC versus the CCD-x one. These relations confirm that the TDI corrector works very well, correcting for the star-trail curvature effect and differential speed, thus transforming the hyperbolic stellar trajectories in the ILMT focal plane into straight lines.


\begin{acknowledgments}
We thank the referee for providing useful comments that have improved the quality of this manuscript. The 4m International Liquid Mirror Telescope (ILMT) project results from a collaboration between the Institute of Astrophysics and Geophysics (University of Li\`{e}ge, Belgium), the Universities of British Columbia, Laval, Montreal, Toronto, Victoria and York University, and Aryabhatta Research Institute of observational sciencES (ARIES, India). The authors thank Hitesh Kumar, Himanshu Rawat, Khushal Singh and other observing staff for their assistance at the 4m ILMT.  The team acknowledges the contributions of ARIES's past and present scientific, engineering and administrative members in the realisation of the ILMT project. JS wishes to thank Service Public Wallonie, F.R.S.-FNRS (Belgium) and the University of Li\`{e}ge, Belgium for funding the construction of the ILMT. PH acknowledges financial support from the Natural Sciences and Engineering Research Council of Canada, RGPIN-2019-04369. PH and JS thank ARIES for hospitality during their visits to Devasthal. B.A. acknowledges the Council of Scientific $\&$ Industrial Research (CSIR) fellowship award (09/948(0005)/2020-EMR-I) for this work. M.D. acknowledges Innovation in Science Pursuit for Inspired Research (INSPIRE) fellowship award (DST/INSPIRE Fellowship/2020/IF200251) for this work. T.A. thanks Ministry of Higher Education, Science and Innovations of Uzbekistan (grant FZ-20200929344).
\end{acknowledgments}

\begin{furtherinformation}

\begin{orcids}
\orcid{0000-0001-5824-1040}{Vibhore}{Negi}
\orcid{0000-0002-7005-1976}{Jean} {Surdej} 
\end{orcids}

\begin{authorcontributions}
This work results from a long-term collaboration to which all authors have made significant
contributions.
\end{authorcontributions}

\begin{conflictsofinterest}
The authors declare no conflict of interest.
\end{conflictsofinterest}

\end{furtherinformation}

\bibliographystyle{bullsrsl-en}

\bibliography{S11-P08_NegiV}

\end{document}